\documentstyle[12pt,psfig]{article}
\begin{document}
\newcommand{\newc}{\newcommand}
\newc{\ra}{\rightarrow}
\newc{\lra}{\leftrightarrow}
\newc{\beq}{\begin{equation}}
\newc{\eeq}{\end{equation}}
\newc{\barr}{\begin{eqnarray}}
\newc{\err}{\end{eqnarray}}
\newc{\eps}{\epsilon}
\newc{\nn}{\nonumber}
\newcommand{\gl}{\lambda}
\newcommand{\PR}{{\it Phys.Rev.}}
\newcommand{\PRL}{{\it Phys.Rev.Lett.}}
\newcommand{\PL}{{\it Phys.Lett.}}
\newcommand{\NP}{{\it Nucl.Phys.}}


~\hfill NTUA  64-97\\
\vspace{1 cm}

\begin{center}
{\bf
 Variational Wave Functionals 
in Quantum Field Theory
}

\vspace*{.5cm}
{\bf
George Tiktopoulos$^{(1)}$
}\\
\vspace*{.5cm}
{\it Physics Department, National Technical University}\\
{\it GR-157 80 Zografou, Athens, Greece}\\
\end{center}

\vspace*{1.5cm}
\begin{abstract}
Variational (Rayleigh-Ritz) methods are applied to local quantum field
theory. For scalar theories the wave functional
is parametrized in the form of a {\it superposition} of Gaussians
and the expectation value of the Hamiltonian is 
expressed in a form that can be minimized numerically.
A scheme of successive refinements of the superposition
is proposed that may converge to the exact functional.
As an illustration, a simple numerical approximation for 
the effective potential is worked out based on minimization
with respect to five variational parameters.
A variational principle is formulated  
for the fermion vacuum energy as a
functional of the scalar fields to which the 
fermions are coupled. The discussion in this 
paper is given for scalar and fermion interactions 
in 1+1 dimensions. The extension to higher dimensions
encounters a more involved structure of ultraviolet 
divergences and is deferred to future work.  
\end{abstract} 
\thispagestyle{empty}
\vfill
\vspace{.5cm}
\hrule
\vspace{.3cm}
{\small
\noindent
$^{(1)}$gtikt@central.ntua.gr

\newpage
  
    1.INTRODUCTION
    
  The variational or Rayleigh-Ritz method is an important
nonperturbative tool in quantum mechanics. In its simplest
form (see e.g. ref.
\cite{sc}
it approximates the ground state 
energy $E_0$  of a Hamiltonian $H$ by the expectation value
\beq
<\Psi \left| H\right| \Psi >/<\Psi \mid \Psi >
\eeq
\noindent
where $\Psi$ is a trial wave function depending on a set
of parameters. The minimum of $E_{\Psi}$ under variation 
of these parameters is the approximation to $E_0$.

  In general, one does not know how close to $E_0$ the 
variational approximation is. It all depends on guessing
the form of the trial function largely on the basis of
intuition. It seems like an impossible task for systems
with many or an infinite number of degrees of freedom
like a quantum field theory (QFT).

  An obvious difficulty with QFT's is that the 
functional integrations required to compute the matrix
elements in (1) are impossible unless $\Psi$ belongs
to a very limited class of functionals essentially
that of Gaussians-times-polynomials in the field 
variables. Moreover,
with a polynomial multiplying the Gaussian, 
one obtains for each of the matrix elements 
a sum of terms proportional to
different powers of the volume resulting in a
trivial thermodynamic limit.

  Still another obstacle are the ultraviolet (UV)
divergences ($E_0$ is UV-divergent even for free
field theories) which require the introduction
of a momentum cutoff and of appropriate cutoff-dependent
counterterms in the Hamiltonian. One then needs to
compute the divergent parts of the counterterms 
exactly -- for instance the exact divergent part of
$E_0$ must be found, otherwise the calculated value 
will be off by an infinite amount.

Early applications of the variational approach \cite{early} have
been largely based on a Gaussian wave functional. Of 
particular interest have been
calculations of the effective potential \cite{effpot} $V_{eff}(v)$
defined as the minimum of the expectation value of $H$ 
under the constraint that the expectation value of the
relevant scalar fiald be $v$.

 Various ideas have been proposed which go beyond the
simple Gaussian wave functional such as the use of an
appropriately unitarily transformed Gaussian \cite{unitary} and
a two-particle-point-irreducible loop expansion of an 
effective action for local composite operators \cite{comp}.

  A different but, nevertheless, also promising approach is 
variational perturbation theory \cite{varpert}, which consists in
making the separation of the Hamiltonian into free
and interaction parts depend on a set of parameters and
improving the perturbation expansion by demanding 
stationarity with respect to these parameters up to a
given order. However, variational perturbation
theory is not really based on a Rayleigh-type
principle, so e.g. the effective potential calculated
by it, cannot be claimed to be an upper bound to the 
exact effective potential.This is true also for a related 
approach employed by the authors of ref. \cite{italians}.
  
   The present work is another attempt to go beyond the simple 
Gaussian functional. Its basic plan is
to parametrize the wave functional by a set of 
parameters as a {\it superposition}
of Gaussians and then to minimize the expectation
value of the Hamiltonian numerically with respect 
to those parameters.

Provided the issues mentioned above, namely the 
thermodynamic limit and the UV divergences,
are successfully met, present computing capabilities 
warrant the hope that minimization with respect to a 
sufficiently large number of parameters can be 
managed and will produce useful results.

 In this paper we explore these possibilities in  
the framework of a local relativistic theory 
in 1+1 dimensions involving a real scalar 
field ${\phi}(x)$ and a fermion field ${\psi}(x)$ with 
Lagrangian density
\beq
L=L_\varphi +L_\psi 
\eeq
\beq
L_\varphi =\frac 12(\frac{\partial \varphi }{\partial t})^2-\frac 12(\frac{%
\partial \varphi }{\partial x})^2-\frac 12m^2\varphi ^2-\frac \lambda
{4!}\varphi ^4
\eeq
\beq
L_\psi =\overline{\psi }(i\not{\partial} -m-\Gamma \varphi )\psi 
\eeq
 
  The generalization to several fields would present
no new problems. However, the extension of the approach 
presented on this paper to higher dimensions encounters a 
more involved structure of ultraviolet 
divergences and will be taken up in future work.  

 In addition to the effective potential, one other
quantity for which one may hope to obtain a
reasonable approximation in terms of a variationally
determined vacuum functional $\Psi$ is the two-point function
\beq
<\Psi \left| \varphi (x_1)\varphi (x_2)\right| \Psi >
\eeq
for all space time separations $x_1-x_2$ since from its 
values at equal times one can,in principle, obtain 
the weight function of its K\"{a}llen-Lehmann representation \cite{lehmann}.
 
  This is a rather long paper. The following
 outline may be useful to the reader.
 
 \begin{enumerate}
 
 \item Introduction.
   
 \item The form of the wave functional is discussed
     for a purely scalar theory.
     
 \item As a first orientation to the problem in its
     simplest form the effective potential for the 
     $\varphi^4$ theory is calculated by means of 
     a single-Gaussian wave functional.
     
 \item A specific form for the superposition of Gaussians     
     is chosen and the field integrations are carried out.
     
 \item The thermodynamic limit is taken after space
     has been divided into cells of equal size assumed
     to be approximately uncorrelated.A sequence of
     increasingly refined approximations is formulated
     which would arguably converge to the exact result.
     
 \item Removal of the UV divergences of the scalar theory.
   
\item  The approach is illustrated by approximating  the 
      effective potential for the $\lambda \varphi^4$ theory
      through numerical minimization with respect to five parameters.
     
 \item Variational approximations are explored 
     for the fermion vacuum energy as a functional
     of the scalar field $\varphi$ to which the 
     fermions are coupled.
     
 \item Removal of UV divergences from the fermion 
     vacuum energy.  
     
 \item Conclusion and prospects.  
 
\end{enumerate}

\newpage
    3.THE FORM OF THE WAVE FUNCTIONAL.
    
    We consider the theory defined by the Lagrangian density
    $L_{\phi}$ in 1+1 dimensions. Initially, we enclose space
    in a box of length V with periodic boundary conditions
    on the real field  ${\varphi}(x)$. The thermodynamic limit 
    $V{\ra}\infty$ will be taken in Section 5.
    
     The Schr\"{o}dinger picture Hamiltonian is
\beq
H_V=\int_{-V/2}^{V/2}dx\{-\frac 12\frac{\delta ^2}{\delta \varphi (x)^2}%
+\frac 12(\frac{\partial \varphi (x)}{\partial x})^2+\frac 12m_u^2\varphi
^2(x)+\frac \lambda {4!}\varphi ^4(x)\} 
\eeq
\noindent
where $m_u$ is the unrenormalized mass and $\lambda$ is the 
unrenormalized coupling constant. We seek to minimize 
the expression
\beq
E_\Psi =\lim _{V\rightarrow \infty }\frac 1V\frac{<\Psi \left| H_V\right|
\Psi >}{<\Psi |\Psi >}
\eeq      
  We take the trial functional in the form of a superposition 
of Gaussians
\beq
\Psi (\varphi )=\sum_g\rho (g)\exp \{-\frac 14\int dxdy(\varphi
(x)-g(x))K(x,y)(\varphi (y)-g(y))\}
\eeq
\noindent
where $K(x,y)$ is a real, symmetric, positive kernel and the
summation runs over some set of functions to be specified.
Ideally, this should be a complete set in some sense.

 The virtue of form (8) is that it allows us to carry 
out the integrations over $\varphi$ explicitly.

 Superposing Gaussians with different kernels would be 
incompatible with the thermodynamic limit. 
Indeed,let $K_1$ and $K_2$ be two such kernels.
Then $<\Psi \left| H_V\right|\Psi>$ and $<\Psi |\Psi>$ ,
as a result of the $\varphi$ integrations would contain
terms proportional to the determinants of $(K_1+K_2)^{\frac 12}$,
${K_1}^{\frac 12}$ and ${K_2}^{\frac 12}$. Assuming that the
kernels ${K_1}(x,y)$ and ${K_2}(x,y)$ depend only on x-y
(for translation invariance), these determinants may be expressed,
for large $V$, as
\beq
\exp \{-\frac V2\int \frac{dk}{2\pi }\log (\widetilde{K}_1+{\widetilde{K}_2}%
)\},\exp \{-\frac V2\int \frac{dk}{2\pi }\log (%
\widetilde{K_1})\}
\eeq
\beq
\exp \{-\frac V2\int \frac{dk}{2\pi }\log (\widetilde{K_2})
\eeq
\noindent
where ${\widetilde K}_1$ and ${\widetilde K}_2$ are the Fourier tranforms
of $K_1$ and $K_2$. But then the requirement that ratios of terms in 
$<\Psi \left| H_V\right|\Psi>$ and $<\Psi |\Psi>$ be finite in
the limit $V{\ra}{\infty}$ leads to
\beq
\int \frac{dk}{2\pi}\log (\frac{\widetilde{K}_1+\widetilde{K}_2}{2\sqrt{%
\widetilde{K}_1\widetilde{K}_2}})=0
\eeq
which implies $K_1=K_2$, since they are positive kernels.

 Similarly, a translationally invariant 
 wave functional consisting of a Gaussian times a polynomial 
consisting of several terms of the form 
\beq
\int dx_1 dx_2 dx_3 \ldots F(x_1-x_2,x_2-x_3, \ldots)%
\varphi(x_1)\varphi(x_2) \ldots
\eeq
\noindent
of {\it different} degrees in $\varphi$ would lead to a sum of 
terms proportional to different powers of V.

 At any rate the single-kernel superposition (8) may be adequate.
One may argue that any "reasonable" functional, 
by analogy to ordinary functions,can be represented 
as a superposition of Gaussians
having the same quadratic term in the exponent.

\newpage

3.THE SINGLE-GAUSSIAN FUNCTIONAL
  
   It is instructive to consider first in some detail the case
 in which the wave functional is a single Gaussian term.
   We consider the wave functional
\beq
\Psi (\varphi ,v)=\exp \{-\frac 14\int dxdy(\varphi (x)-v)K(x-y)(\varphi
(y)-v)\}
\eeq
\noindent
in which translation invariance is explicit and the average value of
the field is arranged to be $v$. The expectation value of the 
energy density is
\barr
E_\Psi &=&\frac 18D^{-1}(0)+\frac 12(-\frac{\partial ^2}{\partial x^2}%
+m_u^2)D(x)|_{x=0}+\frac 12m_u^2v^2 + \nonumber \\
 & &\frac \lambda {24}v^4+\frac \lambda 4v^2D(0)+\frac \lambda 8D(0)^2
\err
where
\beq
D(x-y)\equiv \frac{<\Psi \left| \varphi (x)\varphi (y)\right| \Psi >}{<\Psi
|\Psi >}-v^2=K^{-1}(x-y)
\eeq
  The variational equation
\beq
\frac{\delta E_\Psi }{\delta D(x)}=0
\eeq
\noindent
can be solved explicitly. We introduce the Fourier transform
\beq
\widetilde{D}(k)=\int dxe^{-ikx}D(x)
\eeq
in terms of which (16) reads
\beq
-\frac 1{8\widetilde{D}(k)}+\frac 12(k^2+m_u^2)+\frac \lambda 4\{v^2+\int 
\frac{dk}{2\pi }\widetilde{D}(k)\}=0
\eeq
 Solving for $\widetilde{D}(k)$ we obtain
\beq
\widetilde{D}(k)=\frac 1{2\sqrt{k^2+m_v^2}}
\eeq
\noindent
where the mass parameter $m_v^2$ (the "renormalized mass") is 
the solution to the "gap equation"
\beq
m_v^2=m_u^2+\frac \lambda 2v^2+\frac \lambda 2\int \frac{dk}{2\pi }%
\widetilde{D}(k)
\eeq
  The integral over $\widetilde{D}(k)$ diverges at large momenta k, so
a cutoff $\Lambda$ must be introduced. For $m_v^2$ to be finite, 
$m_u^2$ must depend on $\Lambda$ so as to cancel the divergent part
of the integral in (18). With no loss of generality we set
\beq
m_u^2+\frac \lambda 2\int_{-\Lambda }^\Lambda \frac{dk}{%
2\pi }\frac 1{2 \sqrt{k^2+M^2}}=M^2
\eeq
 The finite mass $M$ replaces $m_u$ as one of the basic constants 
 of the theory, the other one being $\lambda$. The gap equation
  becomes
\beq
m_v^2=M^2+\frac 12\lambda v^2+\frac 1{8\pi }\lambda \log (\frac{M^2}{m_v^2})
\eeq
  The minimum of $E_{\Psi}$ is
\barr
E_\Psi& =&\frac 12\int_{-\Lambda }^\Lambda \frac{dk}{2\pi }|k|-\frac
1{2\lambda} (M^2-\frac \lambda 2\int_{-\Lambda }^\Lambda \frac{dk}{2\pi }\frac 1{2\sqrt{%
k^2+M^2}})^2+\frac 1{8\pi }m_v^2+\frac \lambda
{24}v^4+ \nonumber \\
& & \frac 12v^2(M^2+\frac \lambda {8\pi }\log (\frac{M^2}{m_v^2})+\frac
1{2\lambda }(M^2+\frac \lambda {8\pi }\log (\frac{M^2}{m_v^2}))^2
\err
The sum of the first two terms can be shown (by calculating
the contribution of the relevant perturbation diagrams) 
to contain the exact UV-divergent part of the energy density
and is independent if $v$. It is interesting to
compare this divergent part to that of the one-loop approximation
to $E_{\Psi}$ which is contained in the zero-point energy
\beq
E_{\Psi,1 loop}=\frac 12 \int \frac {dk}{2 \pi} \sqrt{k^2+m_v^2}+\ldots
\eeq
  Clearly, the variational energy is "infinitely" lower.
    
 We may define the finite expression
\beq
{V_{eff}}^{(1)}(\lambda ,M,v)=E_\Psi (\lambda ,M,v,\Lambda )-
E_\Psi (\lambda ,M,0,\Lambda)
\eeq
\noindent
as the approximate effective potential. It is an even function of $v$
and it increases monotonically with $v^2$ as long as
\beq
\frac \lambda {8\pi M^2}<\min_x (\frac{2+x}{2\log x})=2.160
\eeq
  For $\lambda/8\pi M^2>2.160$ the approximate effective potential 
develops a minimum at some nonzero value $v_0^2$ of $v^2$. This minimum
value becomes negative for $\lambda/8\pi M^2>2.439 $  indicating
vacuum states that break the $\varphi \ra -\varphi$ symmetry
of the Hamiltonian.

 It is worth pointing out that for $\lambda/8\pi M^2>2.439 $ 
(symmetry breaking case) a lower value for the energy density 
is achieved by superposing just two Gaussians, 
one centered at $\varphi=v_0$ and the other at $\varphi=-v_0$:
\barr
\Psi (\varphi )&=&\cos \beta \Psi _{+}+\sin \beta \Psi _{-} \\
\Psi _{\pm }(\varphi )&=&\Psi _{\pm }(\varphi ,\pm v_0)  \nonumber
\err
  Note that in the $V \ra 0$ limit the matrix elements
\beq
<\Psi _{+}|\Psi _{-}>,<\Psi _{+}|\varphi |\Psi _{-}>,
<\Psi_{+}|H|\Psi_{-}>    
\eeq
\noindent
vanish, because they are proportional to
\beq
\exp \{-\frac 12v_0^2\int dxdyK(x-y)\}.
\eeq
 It follows that as $\beta$ ranges from $0$ to $\pi/2$ the expectation
 value of $\varphi$ varies from $v_0$ to $-v_0$ while $E_{\Psi}$
 remains  constant and equal to $V_{eff}^{(1)} (\pm v_0)$.
 Note that this "improved" effective potential is a convex 
 function of $v$ as expected for the exact effective potential.

\newpage
 4.SUPERPOSITION
 
 In this Section we discuss the general superposition (8). We make a 
 special choice for the set of g functions and we obtain a convenient
 expression for the expectation value $E_{\Psi}$ so that the limit
 $V \ra \infty$ can be taken in the next Section.
 
  We begin by writing the kernel $K$ as
  \beq
 K=(AA^{\dagger })^{-1} 
 \eeq
 where $A$ is a nonsingular real integral kernel to be specified.

 We make the summation over $g$ in Eq.(8) concrete by expressing 
 $g(x)$ in terms of a real constant $v$ and a set of
 parameters $c_1,c_2,...$ 
 \beq
 g(x)=v+\sum_\alpha c_\alpha Ah_\alpha (x) 
 \eeq 
 where $\{h_{\alpha}(x) \}$ is an orthonormal set of
 functions in ${\cal L}_2(-\infty,\infty)$. We have set
 \beq
 Ah_\alpha (x) \equiv \int dyA(x,y)h_\alpha (y)
 \eeq
 
   Ideally, the set $\{h_{\alpha}(x) \}$ should be complete. 
 Realistically,though, we shall have to assume that even 
 with an incomplete set of such basis functions good results
 will be achieved provided their linear combinations
 can adequately represent functions that are localized
 anywhere in (configuration) space and also lie,
 in Fourier space, within the range of wave numbers
 characteristic of the field-theoretic system
 being discussed.
 
  The summation over $g$ in Eq.(8) will be realized
  as an integration over the $c$ parameters
  \beq
 \sum_g\rho (g)\rightarrow \int (dc)\rho (c)
 \eeq
  where we introduced the shorthand notation
  \barr
 (dc)&=&\prod_\alpha dRec_\alpha dImc_\alpha \\
 \rho (c)&=&\rho (c_1,c_2,...) 
 \err
 
  With respect to the set of functions $\{h_{\alpha}(x) \}$ 
  the essential choice is the subspace they
  span, since within the same subspace any change of
  basis is equivalent to a linear unitary transformation 
  of the set of $c$ variables.
  
   Carrying out the $\varphi$ integrations we obtain
   \barr
 <\Psi |\Psi >&=&(\det K)^{-\frac 12}\int (dc)(dc^{\prime })\rho (c)\rho^{*}
(c^{\prime })\exp \{-\frac 18\sum_\alpha (c_\alpha -c_\alpha ^{*\prime })^2\} \\
 <\Psi |\varphi |\Psi >&=&(\det K)^{-\frac 12}\int (dc)(dc^{\prime })\rho
(c)\rho^{*} (c^{\prime })\exp \{-\frac 18\sum_\alpha (c_\alpha -c_\alpha
^{*\prime })^2\} \nn \\
& &\sum_\beta Ah_\beta (x)(\frac{c_\beta +c_\beta ^{*\prime }}2)
\err

  Note that the $c$ variables are "decoupled"
  in the exponential of these expressions and the same
  is true for analogous expressions for the expectation
  value of the product of any number of fields. This
  is the reason we have expanded $g(x)$ in terms of the set
  $\{Ah_{\alpha} \}$ rather than $\{h_{\alpha} \}$.
  
   In what follows we shall use the shorthand notation
  \beq
  <f(c,c^{\prime})>=\frac{\int (dc)(dc^{\prime })\rho
(c)\rho^{*} (c^{\prime })f(c,c^{\prime})\exp \{-\frac
18\sum_\alpha (c_\alpha -c_\alpha ^{\prime *})^2\}}{\,\,\,\,\,\,\,\int
(dc)(dc^{\prime })\rho (c)\rho^{*}(c^{\prime })\exp \{-\frac 18\sum_\alpha
(c_\alpha -c_\alpha ^{\prime *})^2\}}
\eeq

The condition
  \beq
 \frac{<\Psi |\varphi (x)|\Psi >}{<\Psi |\Psi >}=v
 \eeq
 imposes on $\rho (c)$ the constraint
 \beq
 <c_\alpha +c_\alpha ^{\prime *}>=0
 \eeq
  
  We introduce the following real, symmetric 
  two-, three- and four-point correlation coefficients
\barr
J_{\alpha _1\alpha _2}^{(\pm )}&=&<(\frac{c \pm c^{\prime *}}2)_{\alpha _1}(\frac{%
c \pm c^{\prime *}}2)_{\alpha _2}> \\
J_{\alpha _1\alpha _2\alpha _3}&=&<(\frac{c+c^{\prime *}}2)_{\alpha _1}(\frac{%
c+c^{\prime *}}2)_{\alpha _2}(\frac{c+c^{\prime *}}2)_{\alpha _3}> \\
J_{\alpha _1\alpha _2\alpha _3\alpha _4}&=&<(\frac{c+c^{\prime *}}2)_{\alpha
_1}(\frac{c+c^{\prime *}}2)_{\alpha _2}(\frac{c+c^{\prime *}}2)_{\alpha _3}(%
\frac{c+c^{\prime *}}2)_{\alpha _4}>
 \err
\noindent
and the real symmetric kernels
\beq
Q^{(\pm )}(x,y)=\sum_{\alpha _1\alpha _2}h_{\alpha _1}(x)J_{\alpha _1\alpha
_2}^{(\pm )}h_{\alpha _2}(y)
\eeq

 We also introduce the {\it connected} four point correlation 
 coefficient
 \beq
J_{\alpha _1\alpha _2\alpha _3\alpha _4}^c=J_{\alpha _1\alpha _2\alpha
_3\alpha _4}-J_{\alpha _1\alpha _2}^{(+)}J_{\alpha _3\alpha _4}^{(+)}
-J_{\alpha _1\alpha _3}^{(+)}J_{\alpha _2\alpha _4}^{(+)}-J_{\alpha
_1\alpha _4}^{(+)}J_{\alpha _2\alpha _3}^{(+)}
\eeq
   It vanishes whenever a subset of its indices is associated with
  c variables uncorrelated to the c variables associated with the
  remaining indices. Note that due to Eq.(40) $J_{\alpha_1 \alpha_2}
  ^{(\pm)}$ and $J_{\alpha_1 \alpha_2 \alpha_3}$ are already
  connected in this sense.
   
   The expectation values needed to calculate $E_{\Psi}$ can now 
   be expressed in terms of these connected correlation 
   coefficients. For the product of two fields we find
   \beq
   \frac{<\Psi |\varphi(x_1)\varphi(x_2)|\Psi >}{<\Psi |\Psi >}=v^2+%
   D(x_1,x_2)
   \eeq
   where D is the real kernel
   \beq
  D=A(1+Q^{(+)})A^{\dagger}
  \eeq
  
    We proceed to satisfy this operator relation by choosing for
 A the real kernel
  \beq
  A=D^{\frac 12}(1+Q^{(+)})^{-\frac 12}
  \eeq
    
    The kernel $D(x_1,x_2)$ represents an approximation
  to the (unrenormalized) two-point function at equal times.
  [Note that knowledge of the two-point function
  at equal times is sufficient to determine the weight
  function of its Kallen-Lehman[8] representation and thus to
   determine it for all times]. We further find 
\beq
  \frac{<\delta \Psi / \delta \varphi (x)|\delta \Psi /\delta
\varphi (x)>}{<\Psi |\Psi >}=\frac 14(A^{\dagger -1}(1-Q^{(-)})A^{-1})(x,x) \\
\eeq                                    
\barr
\frac{<\Psi |\varphi ^4(x)|\Psi >}{<\Psi |\Psi >}%
&=&3(v^2+D(x,x))^2-2v^4-3(AQ^{(+)}A^{\dagger })(x,x)^2+ \nn \\
    & &4v\sum_{\alpha _1\alpha _2\alpha _3}J_{\alpha _1\alpha _2\alpha_3}
Ah_{\alpha _1}Ah_{\alpha _2}Ah_{\alpha 3}+ \nn \\
& &\sum_{\alpha _1\alpha _2\alpha _3\alpha _4}J_{\alpha _1\alpha _2\alpha_3
\alpha _4}Ah_{\alpha _1}Ah_{\alpha _2}Ah_{\alpha 3}Ah_{\alpha _4} 
\err
  Using Eqs.(46), (47), (48), (49) and (50) we arrive at the
  following expression for the expectation value of the
  energy density
\barr
E_\Psi &=&\frac 12\int \frac{dx}V\{\partial _x\partial
_yD(x,y)|_{y=x}+m_u^2D(x,x)\}+\frac 12m_u^2 v^2+\frac 18\int \frac{dx}%
VD^{-1}(x,x)+ \nn \\
& &\frac \lambda {24}\{v^4+3\int \frac{dx}VD(x,x)^2+6v^2\int \frac{dx}VD(x,x)\} \nn \\
& &\frac 18\int \frac{dx}V\{D^{-1}[(1+Q^{(+)})^{\frac
12}(1-Q^{(-)})(1+Q^{(+)})^{\frac 12}-1]\}(x,x)+ \nn \\
& &\frac{\lambda}{24}\{4v\int \frac{dx}V\sum_{\alpha _1\alpha _2\alpha _3}
J_{\alpha _1\alpha_2
\alpha _3}Ah_{\alpha _1}(x)Ah_{\alpha _2}(x)Ah_{\alpha 3}(x)+ \nn \\
& &\int \frac{dx}V\sum_{\alpha _1\alpha _2\alpha _3\alpha _4}J_{\alpha
_1\alpha _2\alpha _3\alpha _4}^cAh_{\alpha _1}(x)Ah_{\alpha _2}(x)Ah_{\alpha
3}(x)Ah_{\alpha _4}(x)\} 
\err

 The expectation value $E_{\Psi}$ is to be minimized with respect to
 $D$ and $\rho$ and the subspace of functions spanned by the
 set $\{h_{\alpha} (x)\}$.
  Those are the "variational parameters".
 \newpage

 5. THE THERMODYNAMIC LIMIT
 
  In this Section we introduce a factorized form for the weight 
  function $\rho (c)$ which allows us to take the limit
  $V \ra \infty$ in our expression for $E_{\Psi}$.

  Assuming translation invariance for D we set
\beq
 D(x,y)\rightarrow D(x-y)
\eeq
  Then the x integrations in all but the last three terms
in (51) are trivial and the $\frac 1V$ factors are removed.

  To remove the $V$ dependence from the last three
terms in (51) we rely on the connectedness of 
the correlation coefficients $J_{\alpha _1\alpha_2}^{(\pm)},
J_{\alpha _1\alpha _2\alpha _3},
J_{\alpha _1\alpha _2\alpha _3\alpha _4}^c$ i.e. the fact 
that they vanish for uncorrelated sets of
indices. More concretely, we assume that $\rho (c)$
factorizes
\beq
\rho (c)=\mu _1(c)\mu _2(c)...\mu _N(c)
\eeq
where $\mu_1,\mu_2 ...\mu_N$ depend on disjoint subsets 
$s_1,s_2,...,s_N$ of the set of $c$ variables. Then
the connected correlation coefficients vanish unless
all their indices belong to only one of the sets 
$s_1,s_2,...,s_N$. Accordingly, it is convenient to label the 
$c$ variables with two indices denoting by
$c_{n,\nu}$ the $\nu$-th variable belonging to
the set $s_n$.

 We proceed to divide space into N cells of equal
length $a$ $(V=Na)$ and to make the special choice
\beq
h_{n,\nu }(x)=h_\nu (x-na)
\eeq
associating the set $s_n$ with the $n$-th cell. We
make the cells equivalent by using the same function
$\mu$ for all the factors in the rhs of (53).

  The connectedness of the correlation coefficients implies 
that we may write
\barr
J_{n_1\nu _1,n_2\nu _2}^{(\pm )}&=&\delta _{n_1,n_2}J_{\nu _1\nu _2}^{(\pm )}\\
J_{n_1\nu _1,n_2\nu _2,n_3\nu _3}&=&\delta _{n_1,n_2}\delta _{n_2,n_3}J_{\nu
_1,\nu _2,\nu _3} \\
J^c_{n_1\nu _1,n_2\nu _2,n_3\nu _3,n_4\nu _4}&=&\delta _{n_1,n_2}\delta
_{n_2,n_3}\delta _{n_3,n_4}J^c_{\nu _1,\nu _2,\nu _3,\nu _4} 
\err
   Each of the last three integrals in (51) splits
into N integrals which are shown to be identical by a shift
 of the integration variable by a multiple of $a$. If we now
 let $N$ and $V$ go to infinity with fixed $a$, 
 the volume $V$ disappears from our expression 
 the factors of $N/V$ being replaced by $1/a$:
\barr 
E_\Psi &=&\frac 12(-\partial ^2+m_u^2)D(x)|_{x=0}+\frac 12m_u^2v^2+ \nn\\
& &\frac \lambda {24}\{v^4+3D(0)^2+6v^2D(0)\}+\frac 18D^{-1}(0)+ \nn \\
& &\frac 1{8a}\sum_{\nu _1,\nu _2}\int dxdyD^{-1}(x-y)h_{\nu
_1}(x) \nn \\
& &[1-(1+J^{(+)})^{\frac 12}(1-J^{(-)})(1+J^{(+)})^{\frac 12}]_{\nu
_1,\nu _2}h_{\nu _2}(y)+ \nn \\
& &\frac \lambda {6a}v\sum_{\nu _1,\nu _2,\nu _3}J_{\nu _1,\nu _2,\nu _3}\int
dxAh_{\nu _1}(x)Ah_{\nu _2}(x)Ah_{\nu _3}(x)+ \nn \\
& &\frac \lambda {24a}\sum_{\nu _1,\nu _2,\nu _3,\nu _4}J^c_{\nu _1,\nu _2,
\nu_3,\nu _4}\int dxAh_{\nu _1}(x)Ah_{\nu _2}(x)Ah_{\nu _3}(x)Ah_{\nu _4}(x) 
\err

  This expression is to be minimized under variations of
$D$, $\mu (c)$ and $a$. Since the lack of
correlation between cells introduced by the factorized
form of $\rho (c)$ is an artificial constraint, we expect
$a=\infty$ at the minimum. However, in actual numerical
calculations, only a finite number of cell basis
 functions can be used. In other words, $\mu (c)$ will
 be of the form
\beq
\mu (c)\rightarrow \mu (c_1,c_2,...,c_K)\Pi _{\nu =K+1}^\infty \delta
(Rec_\nu )\delta(Imc_\nu)
\eeq
  Thus the value of $a$ at the minimum will 
be finite and the approximate vacuum functional 
will not be translationally invariant. It will only be 
invariant under translations by multiples of $a$.
 
 Presumably, as the approximation gets refined with K (= number of basis
functions) increasing to infinity, the cell size $a$ will also
grow to infinity. At any rate a value of $a$ significantly
larger than any characteristic length in the theory
would signal that the effect of the assumed lack of correlation
between cells (an "edge effect") is negligible. 
It would be evidence of a good approximation.
 
  An obvious choice for the set of basis functions $\{h_{\nu} (x)\}$
 is any complete orthonormal set of functions in ${\cal L}_2 (-a/2,a/2)$
 which are defined to be zero outside the interval $(-a/2,a/2)$,
 since then the orthogonality between $h_{\nu_1} (x-n_1 a)$ and
 $h_{\nu_2} (x-n_2 a)$ for $n_1 \neq n_2$ would be trivially 
 satisfied. An example is the familiar set of particle-in-a-box 
 eigenfunctions
  
\barr
 h_{\nu} (x)&=&\sqrt{\frac 2a}\sin\{\nu \pi (\frac xa+\frac 12)\},\;\;\; 
 x\epsilon (-a/2,a/2) \nn  \\
 &=&0 ,\;\;\;\;\;\;\;\;\;\;\;\;\;\;\;\;\;\;\;\;\;\;\;\;\;\;\;\;\;\;\;
  x \not \epsilon (-a/2,a/2)                                   
\err
\beq
\eeq
  Linear superpositions of such functions or their derivatives,
  in general, are discontinuous at cell boundaries. 
 An alternative, attractive choice are {\it wavelets} \cite{chui}. These 
 are of the form
 \barr
 \psi _{n,j}(x)&=&2^{\frac j2}a^{-\frac 12}\psi (2^j\frac xa-n) \\ 
            n,j&=&0,\pm 1,\pm 2,...
 \err
 where $\psi$ is specially constructed to ensure completeness
 and orthonormality
\beq
\int_{-\infty }^\infty dx\psi _{n,j}^{*}(x)\psi _{n^{\prime },j^{\prime
}}(x)=\delta _{n,n^{\prime }}\delta _{j,j^{\prime }}
\eeq

  The "mother wavelet" $\psi$ can be chosen to possess
 continuous derivatives up to any given order.
 
 If $\psi$ is centered at $x=0$ with a width of order $1$,
 then $\psi_{n,j}$ is centered at $x=na/2^j$ with a width
 of order $a/2^j$, while its Fourier-space width is $\sim 2^j/a$.
 Thus, in order to use wavelets in the present context,
 we would replace the index $\nu$  by a set of two integers
 \beq
 \nu \rightarrow \{j,q\} 
 \eeq
 which take values
 \barr
 j&=&0,1,2,3 \ldots \nn \\
 q&=&0,\pm 1,\pm 2,\pm 3,\ldots \nn \\
 -2^{j-1}& <&q \leq 2^{j-1}
 \err
 
  We define
 \barr
 h_{n,\{j,q\}}(x)&=&2^{\frac j2}a^{-\frac 12}\psi (2^j(\frac xa-n)-q) \nn \\
                 n&=&0,\pm 1,\pm 2,\ldots
\err

  Eq.(54) is satisfied because
\beq
h_{n,\{j,q\}}(x)=h_{0,\{j,q\}}(x-na)                 
\eeq 
 Thus $h_{n,(j,q)}$ is centered at $x \sim na+q/2^j a$
with a width of the order of $a/2^j$

  The set of orthonormal functions defined by
 Eqs.(66) and (67) is not complete since
 wavelets with widths greater than $a$ (i.e. those
 corresponding to $j<0$ ) are not included. This
 again is an "edge effect" expected to become
 negligible for large $a$.
 
 \newpage
 
 6. REMOVAL OF UV DIVERGENCES

  In this Section we separate out of $E_{\Psi}$ the
  additive UV-divergent term. We also determine
  the cutoff dependence of the unrenormalized mass
  $m_u$ which renders the theory finite.
   
   The variational equation
   \beq
 \frac{\delta E_\Psi }{\delta D(x)}=0  
 \eeq
  cannot be solved explicitly for $D(x)$ in the general case.
 However, its large $k$ behaviour can be obtained provided
  the Fourier transforms of the basis functions
  $\tilde{h}_{\nu}(k)$ vanish fast enough. Assuming
  that all terms containing $\tilde{h}_{\nu}(k)$
  drop out we obtain for $\widetilde{D}(k)$ the same
  asymptotic behaviour as that of the free theory:
  \beq
  \widetilde{D}(k)_{k\rightarrow \pm \infty }\rightarrow \frac 1{2\left|
k\right| }
\eeq
 This behaviour is indicated for the two-point function 
 in 1+1 and 2+1 dimensions in perturbation theory but not
 in 3+1 dimensions for an interacting theory.
 
  Going back to $E_{\Psi}$ we note that the quantity
\barr
\frac 1{2v}\frac{\partial E_\Psi}{\partial v}&=&\frac 12m_u^2+\frac \lambda
{12}v^2+\frac \lambda 4\int \frac{dk}{2\pi }\widetilde{D}(k)+ \nn \\
& &\frac{\lambda}{12av}
\sum_{\nu _1,\nu _2,\nu _3}J_{\nu _1,\nu _2,\nu _3}\int
dxAh_{\nu _1}(x)Ah_{\nu _2}(x)Ah_{\nu _3}(x) 
\err 
must be finite. Therefore,assuming that the last term
is UV finite, the divergent part of the
integral over $\tilde{D}$ (the "tadpole" term) must be 
cancelled by that of the unrenormalized mass term. Just as 
in the single-Gaussian case, with no loss of generality, we set
\beq
 m_u^2(\Lambda )+\frac \lambda 2\int_{-\Lambda }^\Lambda \frac{dk}{2\pi }%
\frac 1{2\sqrt{k^2+M^2}}=M^2
\eeq
thereby introducing the finite mass parameter M to
take the place of $m_u$.
 
 To proceed, we need to extend Eq.(70) and assume that
 $\tilde{D}$ has an asymptotic expansion of the type
 \barr
 \widetilde{D}(k)_{k\rightarrow \pm \infty }&=&\frac 1{2\left| k\right| }(1+ 
\frac{\gamma _1}{\left| k\right| ^\alpha }+0(\left| k\right| ^{-\beta }))  \\
& &\alpha >0 , \beta > 2  \nn 
\err
where $\gamma_1$, $\alpha$ and $\beta$ are constants. Again
this is supported by perturbation theory in which $\alpha=2$. 
As a consequence of Eq.(73) we find that the expressions
\barr
 & &\int_{-\infty }^\infty \frac{dk}{2\pi }(\widetilde{D}(k)-\frac 1{2\sqrt{%
k^2+M^2}}) \\
 & &\int_{-\infty }^\infty \frac{dk}{2\pi }(\frac 1{8\widetilde{D}(k)}+\frac
12k^2\widetilde{D}(k)-\frac 12\left| k\right|) 
\err
are UV finite. Thus our expression for $E_{\Psi}$ takes the form
\barr
 E_\Psi& =&\int \frac{dk}{2\pi }|k|-\frac 1{2\lambda}
(M^2-\frac \lambda 2\int_{-\Lambda }^\Lambda \frac{dk}{2\pi }\frac 1{2\sqrt{%
k^2+M^2}})^2+ \nn \\
& &\int \frac{dk}{2\pi }(\frac 1{8\widetilde{D%
}(k)}+\frac 12k^2\widetilde{D}(k)-\frac 12\left| k\right| )+\frac{\lambda v^4%
}{24}+ \nn \\
& &\frac{v^2}2(M^2+\frac \lambda 2\int \frac{%
dk}{2\pi }(\widetilde{D}(k)-\frac 1{2\sqrt{k^2+M^2}}))+ \nn \\
& &\frac 1{2\lambda }(M^2+\frac \lambda 2\int 
\frac{dk}{2\pi }(\widetilde{D}(k)-\frac 1{2\sqrt{k^2+M^2}}))^2 
\err
\begin{center}
 +terms containing $h_\nu $
\end{center}

   The sum of the first two terms of this expression contains the
 exact UV-divergent part of the vacuum energy density of the
 theory. These terms are independent of $v$ and of the variational 
 parameters $D(x)$, $\mu (c)$, $a$, and $\{h_{\nu} \}$. We may 
 simply drop them and retain the following finite expression 
 to be minimized.
 \barr
E_{\Psi}&=&\int \frac{dk}{2\pi }%
(\frac 1{8\widetilde{D}(k)}+\frac 12k^2\widetilde{D}(k)-\frac 12\left|
k\right| )+ \nn \\
& &\frac{v^2}2(M^2+\frac \lambda 2\int \frac{dk}{2\pi }(\widetilde{D}(k)-\frac
1{2\sqrt{k^2+M^2}}))+\frac{\lambda v^2}{24}+ \nn \\
& &\frac 1{2\lambda }(M^2+\frac \lambda 2\int \frac{dk}{2\pi }(\widetilde{D}%
(k)-\frac 1{2\sqrt{k^2+M^2}}))^2+ \nn \\
& &\frac 1{8a}\sum_{\nu _1,\nu _2}\int dxdyD^{-1}(x-y)h_{\nu
_1}(x) \nn \\
& &\;\;\;\;[1-(1+J^{(+)})^{\frac 12}(1-J^{(-)})(1+J^{(+)})^{\frac 12}]_{\nu
_1,\nu _2}h_{\nu _2}(y)+ \nn \\
& &\frac \lambda {6a}v\sum_{\nu _1,\nu _2,\nu _3}J_{\nu _1,\nu _2,\nu _3}\int
dxAh_{\nu _1}(x)Ah_{\nu _2}(x)Ah_{\nu _3}(x)+ \nn \\
& &\frac \lambda {24a}\sum_{\nu _1,\nu _2,\nu _3,\nu _4}J^c_{\nu _1,\nu _2,\nu
_3,\nu _4}\int dxAh_{\nu _1}(x)Ah_{\nu _2}(x)Ah_{\nu _3}(x)Ah_{\nu _4}(x)
\err

 In higher-dimensional theories the assumption that the 
 last term in the rhs of \mbox{Eq.(71)} 
 is UV-finite cannot be maintained. This is 
 clear even in the super-renorma\-lizable (2+1)-dimensional 
 case in which the UV-divergent part of $m_u^2$ is calculable. 
 In that model, in addition to the 
 "tadpole" term $\lambda D(0)/2$ associated with 
 the diagram (a) of Fig.1, there is also the divergent 
 contribution of the two-loop diagram (b) of Fig.1 which
 must be related to the last term in Eq.(71).

\begin{picture}(300,200)(0,0)
\centering
\put(105,125){\circle{50}}
\put(255,125){\circle{50}}
\put(55,105){\line(1,0){100}}
\put(205,125){\line(1,0){100}}
\put(100,60){(a)}
\put(250,60){(b)}
\put(40,35){FIG.1 \,\,\,\,\,Diagrams contributing 
to the UV divergence}
\put(80,20){of the two-point function in 2+1 dimensions.}
\end{picture}

 In 3+1 dimensions a host of interrelated UV problems
must be faced. The large-momen\-tum beha\-viour of the connected
 two-point function $D(k)$ is not the same as that of
 the free theory but is de\-termined by the anoma\-lous
 di\-mensions of the field (provided an UV fixed point
 exists); the wave function renorma\-lization constant
 is UV-divergent and so is the unre\-normalized coupling
 $\lambda$. It seems evi\-dent that in these theories, in
 order to achieve the minimum of $E_\Psi$, the 
 corre\-lation coefficients must be such as to make
 the sums in Eqs.(71) and (77) diverge 
 in the $\Lambda \ra \infty$ limit.

\newpage

7.A NUMERICAL CALCULATION OF THE EFFECTIVE POTENTIAL

In this section we calculate  numerically an approximation 
to the effective potential using 
the simplest superposition of Gaussians formed
with just one basis function per cell:
\beq
g(x)=v+\sum_nc_nh(x-na)
\eeq                                        
and with real coefficients $c_n$.
In this case the sum of the last three terms 
in Eq.(77) is simplified to
\barr
& &\frac 1{8a}[1-(1-J^{(+)})(1-J^{(-)})]\int \frac{dk}{2\pi }|\widetilde{h}%
(k)|^2/\widetilde{D}(k)+ \nn \\
& &\frac \lambda {6a}vJ_3\int dx[Ah(x)]^3+\frac \lambda {24a}J_4\int
dx[Ah(x)]^4 
\err
                                                 
where  we have set
\barr
J^{(\pm )}&=&<(\frac{c \pm c^{\prime }}2)^2> \\\
J_3&=&<(\frac{c+c^{\prime }}2)^3> \\
J_4&=&<(\frac{c+c^{\prime }}2)^4>-3<(\frac{c+c^{\prime }}2)^2>^2 \\
A(x-y)&=&D^{1/2}(x-y)/(1+J^{(+)})^{1/2}
\err
                                                
 The function $h(x)$ must satisfy the orthonormality condition
\beq
\int dxh(x-na)h(x-n^{\prime }a)=\delta _{nn^{\prime }}
\eeq

which may be expressed in terms of the Fourier transform 
of $h$ as
\beq
\frac 1a\sum_{s=-\infty }^\infty |\widetilde{h}(k+\frac{2\pi s}a)|^2\equiv 1
\eeq

Subject to this constraint, $h(x)$ is itself a variational
parameter.  However, for simplicity, we make a definite choice.
Note that if we take the Fourier tranform of $h(x)$ to be of the form
\beq
\widetilde{h}(k)=\frac{\widetilde{f}(k)}{(a\sum_{s=-\infty }^\infty |%
\widetilde{f}(k+\frac{2\pi s}a)|^2)^{1/2}}
\eeq
then Eq.(85) is satisfied for any f for which the sum converges
(certain integrability conditions are obviously also  necessary).
 For f(x) we take the second-order cardinal spline and find
\beq
\widetilde{h}(k)=[\frac 2k\sin (\frac k2)]^2[1-\frac 23\sin ^2(\frac
k2)]^{-1/2}
\eeq
 which is the scaling function for the associated wavelet.
 
Since $D(x)$ approximates the (unrenormalized, equal time) two-point
function it is appropriate to adopt the K\"{a}llen-Lehman form for it i.e.
\beq
\widetilde{D}(k)=\int_0^\infty d(\tau ^2)\frac{\sigma (\tau ^2)}{2\sqrt{%
k^2+\tau ^2}}
\eeq
with the positive weight function $\sigma$ to be determined.  For
instance, approximating $\sigma$ by a sum of delta
function terms, we would write
\beq
\widetilde{D}(k)=\sum_j\frac{\gamma _j}{2\sqrt{k^2+\tau ^2}}%
,\,\,\,\,\,\,\,\sum_j\gamma _j=1
\eeq

In order to keep as few variational parameters as possible, we simply
take only one such term
\beq
\widetilde{D}(k)=\frac 1{2\sqrt{k^2+m^2}}
\eeq

Finally, we adopt a Gaussian-times-polynomial form for the weight
function $\mu (c)$ so that the integrations over the $c$'s can
 be explicitly done:
 \beq
\mu (c)=(1+u_1c+u_2c^2+u_3c^3)\exp (-\gamma c^2/2) 
\eeq

The constraint (40) can be used to
express e.g. $u_3$ in terms of $u_1,u_2$ and $\gamma$.
 Thus, for given values of $\lambda$, $M$ and $v$, 
the expectation value of the energy density is an
explicit function of the five variational parameters
$m,a,\gamma,u_1,u_2$.
              
To find the minimum of $E_{\Psi}$ a FORTRAN minimization 
program was run on a PC with a Pentium 133
CPU. Each numerical evaluation of $E_{\Psi}$ 
took $.5$ sec on the average.  Each search for the
minimum $E_S$ of $E_{\Psi}$ took $\sim 2$min.
The product  $ma$ ranged roughly from 0.6 to 1.3, so 
the values found for $E_S$ cannot be claimed to 
be close to those of the exact energy density. Nevertheless, 
they provide a {\it rigorous upper bound} to the corresponding
values of the exact $V_{eff}(v)$. 

In Fig.2 the lowering $E_S-E_G$ of the vacuum energy 
density achieved by the superposition is plotted vs. the
coupling $\lambda$ for $v=0$. Being zero at $\lambda =0$
it decreases steadily with increasing $v$.

In Fig.3 $E_S$ and $E_G$ are plotted vs. $v$ 
for $\lambda/M^2 =50$ and $\lambda/M^2 =100$.
For $\lambda/M^2 =100$ $E_S$ displays a minimum
at $v=v_0 \sim .8$.

Just as in the single-Gaussian case of Section 3, if a linear 
combination of $\Psi (v_0)$ and $\Psi (-v_0)$
were used as a wave functional the energy 
density in the interval $(-v_0,v_0)$ would be lowered 
to the value of $E_S$ at $v=\pm v_0$ and the 
curve would become convex. Presumably, this "extra" 
superposition using different values of $v$ 
would not have been necessary, if a complete 
set of basis functions had been used.

\newpage

\begin{figure}[h]
\centerline{\hbox{\psfig{figure=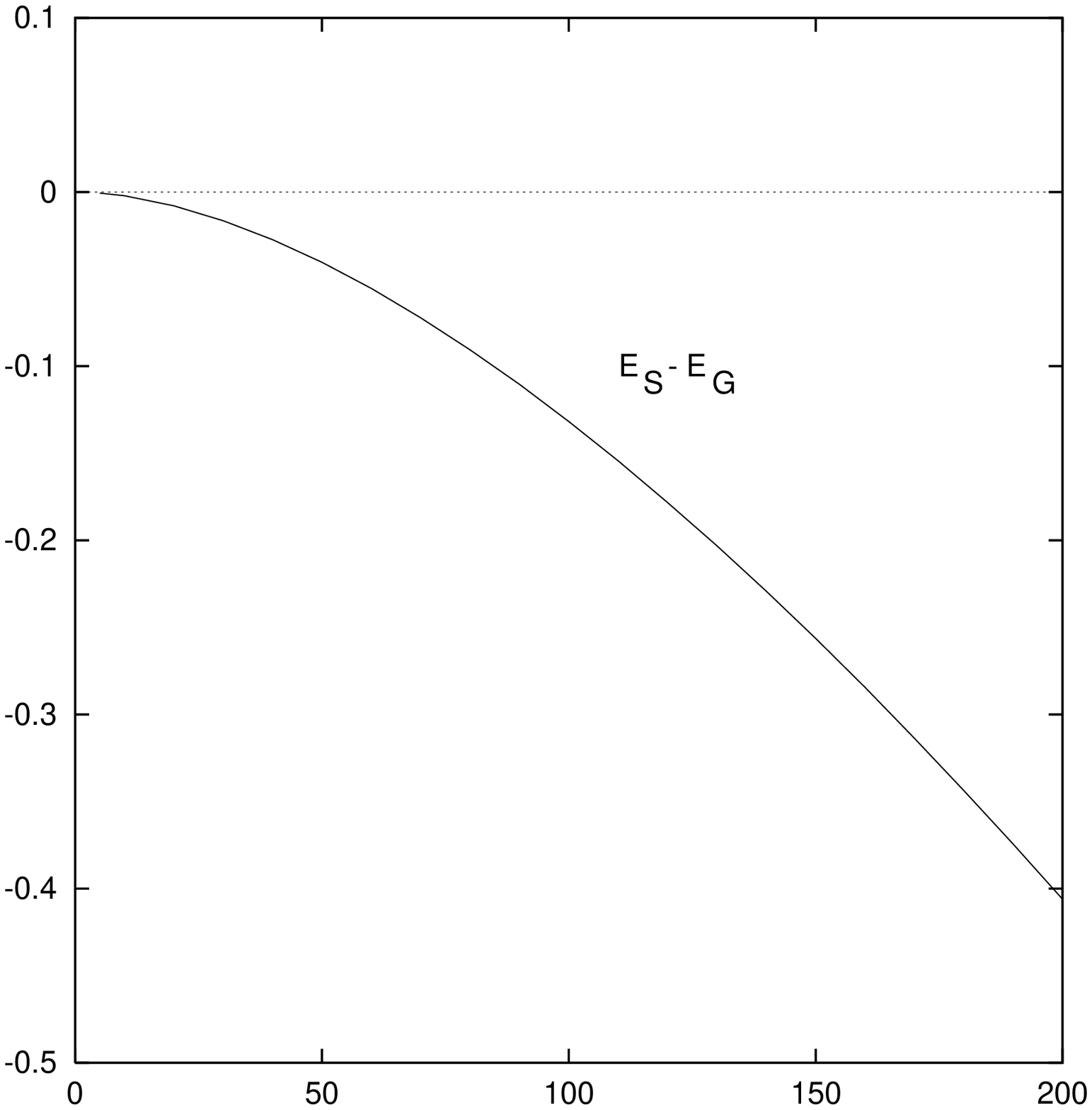,height=6.5cm,angle=-90}}}
\end{figure}
\begin{center}
$\lambda/M^2 \longrightarrow $
\end{center}
\hspace{1in}FIG.2 \parbox[t]{3.5in}
{The improvement $E_S-E_G$ on the approximate vacuum  
energy density, in units of $M^2$, obtained by a superposition 
of Gaussian functionals vs. the coupling $\lambda$ for $v=0$.}
\vspace{12mm}

\begin{figure}[h]
\centerline{\hbox{\psfig{figure=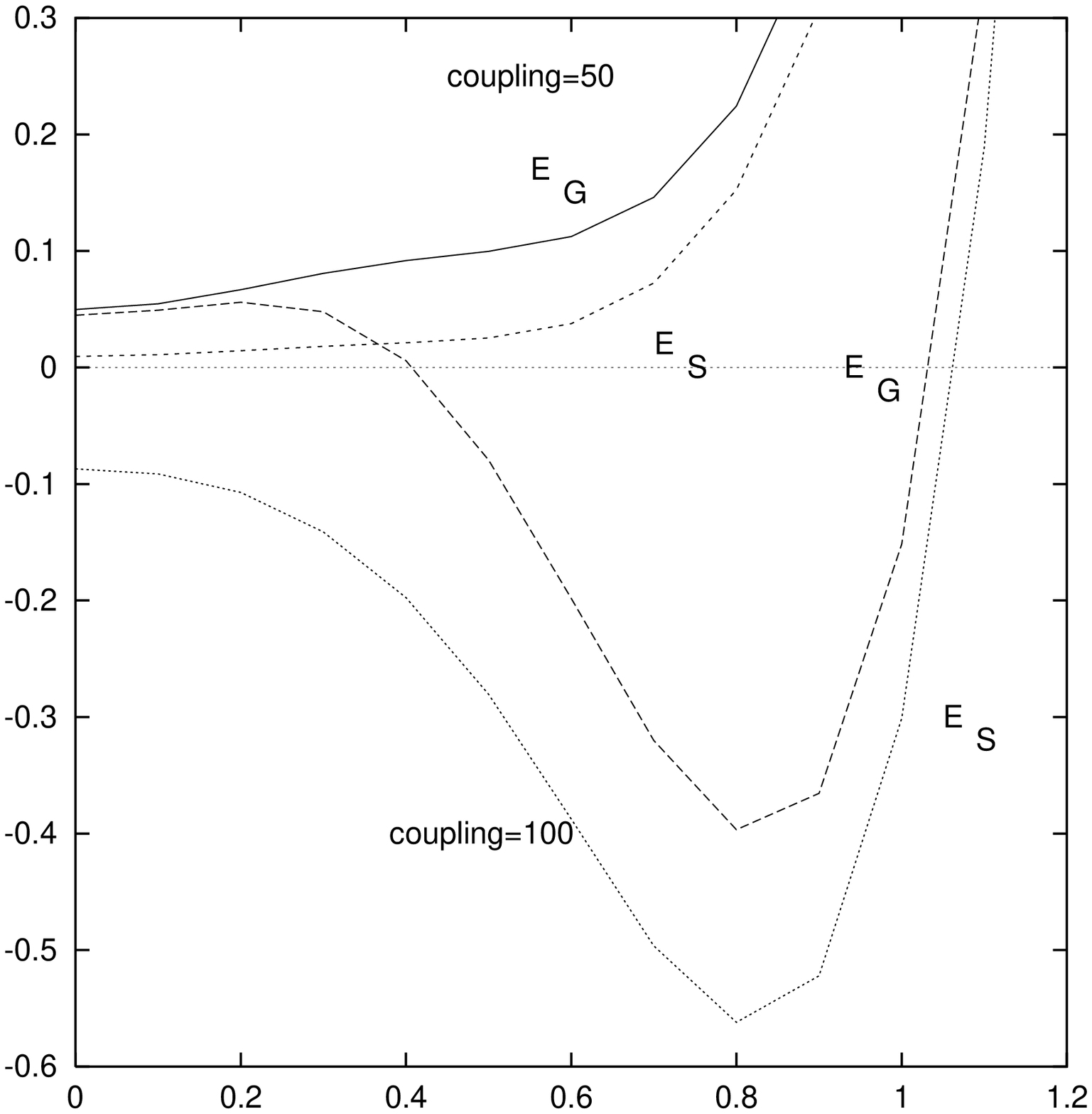,height=6.5cm,angle=-90}}}
\end{figure}
\begin{center}
$v \longrightarrow $
\end{center}
\hspace{1in}FIG.3 \parbox[t]{3in}{Comparison of 
the approximate energy densities $E_S$ and $E_G$, in 
units of $M^2$, plotted against the exp. value $v$ 
of the field for $\lambda/M^2=50$ and $\lambda/M^2 =100$.}
\vspace{12mm}

\newpage

8.FERMIONS

We now consider coupling a fermion field $\psi (x)$, to the
real scalar field $\varphi(x)$ with the Hamiltonian
\beq
H=H_S+H_F
\eeq
where
\barr
H_S&=&\int dx\{-\frac 12\frac{\delta ^2}{\delta \varphi (x)^2}+\frac
12(\frac{\partial \varphi (x)}{\partial x})^2+\frac 12m_u^2\varphi ^2(x)+
\frac \lambda{4!}\varphi ^4(x)\} \\
H_F&=&\int dx\frac 12[\psi ^{\dagger }(x),(-i\alpha \frac \partial {\partial
x}+\beta m+\Gamma \varphi (x))\psi (x)]
\err
The hermitean matrix $\Gamma$ is some linear combination
of $\beta$ and $i \beta \gamma_5$ with real coefficients. 
We deal explicitly with the (1+1)-dimensional theory
(so we could have
replaced $\alpha$, $i\beta \gamma_5$, and $\beta$ by the 
Pauli matrices $\sigma_1, \sigma_2$ and $\sigma_3$)
but, actually, the discussion in this Section is valid
for any dimension (with obvious notational adjustments).  

Let $E_f (\varphi)$ be the ground state energy density of $H_F$ for a given
(c-number) scalar field configuration $\varphi (x)$.  Then the ground state
energy of H is the same as that of the Schr\"{o}dinger-picture operator
\beq
\widehat{H}=H_S+VE_f(\varphi )
\eeq
  In this Section we shall look for variational approximations 
to $E_f (\varphi)$ by explicit functionals of $\varphi$.

 Consider a general Hamiltonian of the form
\beq
H_f=\int dxdx^{\prime }\frac 12[\psi ^{\dagger }(x),h(x,x^{\prime })\psi
(x^{\prime })]
\eeq
where $\psi (x)$ is a fermion field operator satisfying the
anticommutation relations
\barr
\{\psi _\alpha (x),\psi _\beta ^{\dagger }(x^{\prime })\}&=&\delta _{\alpha
,\beta }\delta (x-x^{\prime }) \nn \\
\{\psi _\alpha (x),\psi _\beta (x^{\prime })\}&=&\{\psi _\alpha ^{\dagger
}(x),\psi _\beta ^{\dagger }(x^{\prime })\}=0
\err

The kernel $h(x,x^{\prime})$ may be any hermitean c-number 
kernel and, in particular,
\beq
h(x,x^{\prime })=\delta (x-x^{\prime })(-i\alpha \frac \partial {\partial
x}+\beta m+\Gamma \varphi (x))
\eeq
 
 If $\{ u_{\alpha}(x) \}$ and $\{ v_{\alpha}(x) \}$ are 
complete sets of positive and negative-energy eigenspinors 
of $h(x,x^{\prime})$ and $\{E_{\alpha}\}$, $\{E_{\alpha}^{\prime} \}$ 
the corresponding eigenvalues, we may expand $h$ as
\beq
h=\sum_r(E_rP_r+E_r^{\prime }P_r^{\prime })
\eeq
where $P_{\alpha}$, $P^{\prime}_{\alpha}$ are the projections 
on the corresponding eigenspinors
i.e. the operators with kernels
\barr
P_r(x,x^{\prime })&=&u_r(x)u_r^{\dagger }(x^{\prime }) \nn \\
P_r^{\prime }(x,x^{\prime })&=&v_r(x)v_r^{\dagger }(x^{\prime })
\err

The ground state of $H_f$ is the state $|\Omega _f>$ which is annihilated
by the totality of the operators
\beq
\int dxu_r^{\dagger }(x)\psi (x),\,\,\,\,\,\,\,\,\,\,\,\,\,\,\int dx\psi
^{\dagger }(x)v_r(x)
\eeq

Assuming $<\Omega _f|\Omega _f>=1$ we have
\beq
<\Omega _f|[\psi (x),\psi ^{\dagger }(x^{\prime })]|\Omega
_f>=\sum_rP_r(x,x^{\prime })-\sum_rP_r^{\prime }(x,x^{\prime })
\eeq
and
\barr
<\Omega _f|H_f|\Omega _f>&=&-\frac 12\sum_{r,s}Tr((P_r-P_r^{\prime
})(E_sP_s+E_s^{\prime }P_s^{\prime })) \nn \\
&=&-\frac 12\sum_r(E_r+|E_r^{\prime }|)Tr(1)
\err
where, in $1+1$ dimensions, $Tr(1)=2$.  From this expression
we must eventually identify and extract the UV-divergent parts.

Now let $|\Omega _0>$ be the ground state of the free Hamiltonian
\barr
H_f^{(0)}&=&\int dxdx^{\prime }\frac 12[\psi ^{\dagger
}(x),h^{(0)}(x,x^{\prime })\psi (x^{\prime })] \nn \\
h^{(0)}(x,x^{\prime })&=&\delta (x-x^{\prime })(-i\alpha \frac \partial
{\partial x}+\beta m)
\err
Clearly, $E_f (\varphi)$ must be equal to the minimum value of
\beq
\frac 1V<W\Omega _0|H_f|W\Omega _0>
\eeq
for all possible choices of a unitary operator $W$.

Actually, the minimum value is attained even if 
we limit $W$ to the class of unitary operators under 
which the field $\psi (x)$ transforms linearly i.e.
\beq
W^{\dagger }\psi _\alpha (x)W=\int dx^{\prime }M_{\alpha ,\beta
}(x,x^{\prime })\psi _\beta (x^{\prime })
\eeq
where $M_{\alpha,\beta}(x,x^{\prime})$ is a unitary kernel.
  Indeed, in this case we have
\beq
<W\Omega _0|H_f|W\Omega _0>=-\frac 12\sum_{r,s}Tr(M(P_r^{(0)}-P_r^{(0)\prime
})M^{\dagger }(E_sP_s+E_s^{\prime }P_s^{\prime }))
\eeq
where $P^{(0)}_r$, $P^{(0)\prime}_r$ are 
the eigenspinor projectors of $h^{(0)}$. 

It suffices 
to choose $M$ so that it maps the subspace of positive
(negative) energy eigenspinors of $h$ into those 
of $h^{(0)}$ respectively, i.e.
\barr
M(\sum_rP_r^{(0)})M^{\dagger }&=&\sum_rP_r \nn \\
M(\sum_rP_r^{(0)\prime })M^{\dagger }&=&\sum_rP_r^{\prime }
\err
Then the rhs of Eq.(107) becomes equal to the ground 
state energy of $H_f$ as given by Eq.(103).
Since
\beq
\sum_rP_r^{(0)}(x,x^{\prime })-\sum_rP_r^{(0)\prime }(x,x^{\prime })=\int 
\frac{dp}{2\pi }\exp (ip(x-x^{\prime })\frac{\alpha p+\beta m}{\sqrt{p^2+m^2}%
 }
\eeq
we conclude that $E_f (\varphi)$, the ground state 
energy density of $H_f (\varphi)$, is the minimum 
value of the expresion
\barr
F(\varphi ,M)&=&-\frac 1{2V}\int dxdx^{\prime }dy\sum_{r,s,t,u}(\int \frac{dp%
}{2\pi }\exp (ip(x-x^{\prime })\frac{\alpha p+\beta m}{\sqrt{p^2+m^2}}%
)_{r,s}M_{s,t}^{\dagger }(x^{\prime },y) \nn \\
& &(-i\alpha \frac \partial {\partial x}+\beta m+\Gamma \varphi
(x))_{t,u}M_{u,r}(y,x)
\err
under variations of the unitary kernel M.
\newpage

9. REMOVAL OF FERMION UV DIVERGENCES

 A formal choice of M for which $F(\varphi,M)$ actually 
attains its minimum value is M\"{o}ller's wave matrix given by 
the time-ordered exponential
\beq
S(\varphi )=T\exp (-i\int_{-\infty }^0dt\exp (ih_0t)\Gamma \varphi (x)\exp
(-ih_0t))
\eeq
where $x$ and $h_0$ are operators on the
space of Dirac wave functions.

  Under fairly general conditions $S(\varphi)$ maps 
eigenspinors of $h^{(0)}$ into eigenspinors of $h$. 
Therefore, according to 
the discussion given in the previous Section, we have
\beq
E_f(\varphi )=F(\varphi ,S(\varphi ))
\eeq

\begin{picture}(350,180)(0,0)
\put(50,100){\circle{50}}
\put(137,100){\circle{50}}
\put(225,100){\circle{50}}
\put(315,100){\circle{50}}
\put(18,100){\line(1,0){13}}
\put(106,100){\line(1,0){12}}
\put(156,100){\line(1,0){12}}
\put(213,117){\line(-3,4){13}}
\put(237,117){\line(3,4){13}}
\put(225,80){\line(0,-1){13}}
\put(303,117){\line(-3,4){13}}
\put(327,117){\line(3,4){13}}
\put(303,84){\line(-3,-4){13}}
\put(327,84){\line(3,-4){13}}
\put(60,30){FIG.4\,\,\,\,Diagrams for the fermion vacuum energy.}
\end{picture}

 The familiar vacuum graphs of Fig.4 correspond to 
successive terms in the expansion of
\beq
E_f(\varphi )-E_f(0)
\eeq
in powers of $\varphi$.  The first term of this expansion, 
the term linear in $\varphi$, is given by
(with $\Gamma =\lambda _1\beta +\lambda _2i\beta \gamma _5$)

\beq
-\frac{\lambda _1}2\int \frac{dp}{2\pi }
\frac m{\sqrt{p^2+m^2}}Tr(1)\frac 1V\int dx\varphi (x)
\eeq

 This is a logarithmically divergent "tadpole"
contribution to $E_f (\varphi)$ which can be cancelled by 
the inclusion of a $c(\Lambda)\varphi (x)$ type counterterm 
in the Hamiltonian density.

  The next term in the expansion of $E_f(\varphi )-E_f(0)$, 
is also logarithmically divergent.  Its divergent piece is
\beq
-\frac 12Tr(1)(\lambda _1^2+\lambda _2^2)\frac 1V\int dx\varphi ^2(x)\log
\Lambda 
\eeq
requiring a $\varphi^2 (x)$ counter term (mass term) in the 
Hamiltonian density to cancel it.

  Unfortunately, the M\"{o}ller matrix $S(\varphi)$  is not available 
in a closed form so that, after we substitute  $F(\varphi,S(\varphi))$  
for $E_f (\varphi)$ in Eq.(96), we can carry out the integrations 
over $\varphi$ explicitly in the matrix element
\beq
\int (d\varphi )\Psi ^{\dagger }(\varphi )\widehat{H}(\varphi )\Psi
(\varphi )
\eeq
with $\Psi (\varphi)$ a sum of Gaussian terms.
We must settle for an approximation $F(\varphi,M)$  
with M such that the $  $ integrations are possible.  
One possibility would be an exponential form e.g.
\beq
M(x,x^{\prime })=\delta (x-x^{\prime })\exp \{i\int dyB(x,y)\varphi (y)\}
\eeq
with $B(x,y)$ a hermitean kernel to be determined variationally.
However, this would not reproduce the UV-divergent parts 
of $E_f(\varphi )-E_f(0)$ exactly.  Clearly, 
getting those divergent parts exactly is necessary - otherwise 
our approximate vacuum energy density would differ from the 
exact value by an infinite amount.

 To remedy the situation we note that these UV divergences 
are associated with the large momentum behavior of the 
matrix elements of $\Gamma \varphi (x)$ between momentum eigenstates 
of the free Dirac hamiltonian
\beq
<p^{\prime },\nu ^{\prime }|\Gamma \varphi (x)|p,\nu
>,\,\,\,\,\,\,\,\,\,\,\,\,\,\,\,\,\,\,\,\nu ,\nu ^{\prime }=\pm 1
\eeq
where $|p,\pm>$ denote the positive (negative) energy
eigenstates of $-i\alpha \frac \partial {\partial x}+\beta m$              
with momentum $p$. Thus if we define, as a 
modified version of $\Gamma \varphi (x)$, 
a cut-off kernel $w(\varphi)$ by
\beq
<p^{\prime },\nu ^{\prime }|w(\varphi )|p,\nu >\equiv \theta (\frac
12|p+p^{\prime }|-p_{\max })<p^{\prime },\nu ^{\prime }|\Gamma \varphi
(x)|p,\nu >
\eeq
and construct the time-ordered exponential
\beq
\widetilde{S}(\varphi )=T\exp \{-i\int_{-\infty }^0dt\exp (ih_0t)w(\varphi
)\exp (-ih_0t)\}
\eeq
then $F(\varphi,\widetilde{S}(\varphi))$ can be shown 
to contain the exact UV-divergent part of $F(\varphi,S(\varphi))$, 
namely
\beq
F(\varphi ,\widetilde{S}(\varphi ))-F(\varphi ,S(\varphi ))=finite\geq 0
\eeq

  Furthermore, by taking the momentum parameter $p_{max}$ large enough, 
one can make the expansion of $F(\varphi ,\widetilde{S}(\varphi ))$ 
in powers of $\varphi$ converge fast, so that it may be approximated  by 
just a small number of terms (one must at least 
include, of course, the UV-divergent terms).

  In general, one cannot expect 
$F(\varphi ,\widetilde{S}(\varphi ))$ 
to be a good approximation to $E_f (\varphi)$ 
(although it would still be an upper bound to it).  
However, one can improve on it by considering  
functionals of the form $F(\varphi ,U\widetilde{S}(\varphi ))$ 
where $U$ is a convenient unitary kernel like the 
simple exponential one given by Eq.(117).  
With $U$ being an exponential and $\widetilde{S}(\varphi )$ 
approximated by a polynomial in $\varphi$, the $\varphi$
integrations in (116) could be carried out explicitly.  In the resulting
expression the kernel $B(x,y)$ would play the role of a variational
parameter.

All this can be readily extended to higher dimensions.
There will be, of course, additional UV-divergent terms
in $E_f(\varphi )-E_f(0)$. For instance, in 3+1 dimensions
there is a $\varphi^4 (x)$ term with a logarithmically
divergent coefficient.
\newpage

10. PROSPECTS.
 
 We conclude with a few remarks concerning
 future prospects for the variational approach
proposed in this paper.

  At present, it appears that UV divergences are
 the main obstacle to extending the calculations
 described in this paper to renormalizable
 field theories in higher dimensions.
 Actually, for super-renormalizable theories in
 2+1 dimensions the divergent parts of the 
 counterterms are calculable since they are 
 associated with a finite set of perturbative diagrams.
 In that sense they should be easier to handle. 
 Things are considerably more involved for (3+1)-dimensional
 theories, as indicated in Section 6.
 
   Irrespective of whether these methods will ultimately
  prove applicable to realistic (3+1)-dimensional
  models (including vector fields), they
  could still be used to test other nonperturbative
  methods by comparing with
  their respective results on lower-dimensional
  systems.
 
    On the numerical side, a more ambitious venture than 
  the  example given in Section 7 would require 
 the introduction of several basis functions
 per cell. Let us assume, for example, a polynomial-
 times-Gaussian form for the weight function: 
 $\mu (c)$
 \barr
 \mu (c_1,c_2,...,c_K)&=&(1+\sum_iu_i^{(1)}c_i+
 \sum_{i,j}u_{i,j}^{(2)}c_ic_j+  \nn \\
 & &\sum_{i,j,k}u_{i,j,k}^{(3)}c_ic_jc_k+...)
 \exp \{-\frac 12\sum_s\gamma _sc_s^2\}
  \err
 
 The number of possible u parameters grows
 fast with the number of basis functions $K$.
 Even if we stopped at the cubic terms, there
 would be 55 parameters for $K=5$ and 285 
 parameters for $K=10$. To those we must add
 the $\gamma$'s, $a$ and the parameters necessary
 to represent $D(x)$. Clearly, the number
 of basis functions needed to achieve a 
 useful result is crucial for the feasibility of 
 the calculation; but it can only be
 determined by numerical experimentation.
 
  On the other hand, the virtue of any
  refinement in the description of the
  wave functional will be gauged by the amount
  by which it actually lowers the expectation value.
  This may help develop a "physical intuition"
  just as in variational calculations of atomic and 
  molecular systems. For instance, it
  may turn out that the form (122)
  is not expeditious or that not all of
  the u coefficients in it are of equal importance.
  
\begin{center}
ACKNOWLEDGMENT
\end{center}
This work was partially supported by the Greek
Secretariat for Research and Development under
Contract No.1170.


\newpage

\begin{thebibliography}{99}

 \bibitem{sc}L.I.Schiff: "Quantum Mechanics", 3rd Ed.,McGraw-Hill, New York,1968.
 
 \bibitem{early} L.I.Schiff, Phys.Rev. {\bf 130}, 458(1963);
 
      G.Rosen, Phys.Rev. {\bf 173}, 1632(1968);
      
      J.Kuti, unpublished.
      
      J.M.Cornwall, R.Jackiw and E.Tomboulis, Phys.Rev. {\bf D10}, 2428(1974);
      
      S.J.Chang, Phys.Rev. {\bf D12}, 1071(1975);Phys.Rep. {\bf C 23}, 301(1975);
      
      T.Barnes and G.I.Ghandour, Phys.Rev. {\bf D22}, 924(1980);
      
      P.M.Stevenson, Phys.Rev. {\bf D30}, 1712(1984);{\bf D12}, 1389(1985).
      
      L.Polley and D.Pottinger eds., "Variational calculations in quantum
      field theory", World Scientific, Singapore, 1988.
      
      C.Best and A.Sch\"{a}fer, "Variational description of statistical field
      theories using Daubechies' wavelets", hep-lat/9402012.
      
      I.I.Kogan and A.Kovner, Phys. Rev. {\bf D51}, 1948(1995);{\bf D52}, 
      3719(1995);
      
      W.E.Brown and I.I.Kogan, "Variational approach to the QCD wave 
      functional: calculation of the QCD beta function", hep-th/9705136;

 \bibitem{effpot} S.Coleman and E.Weinberg, Phys.Rev. {\bf D7}, 1888(1973);

      S.Weinberg, Phys.Rev. {\bf D7}, 2887(1973);
      
      R.Jackiw, Phys.Rev. {\bf D9}, 1686(1973);
      

 \bibitem{unitary} L.Polley and U.Ritschel, Phys.Lett. {\bf B221}, 44(1989);

      I.Yotsuyanagi, Z.Phys. {\bf C35}, 453(1987);Phys.Rev. {\bf D39}, 3034(1989);
      
      
 \bibitem{comp} H.Verschelde and M.Coppens, Phys.Lett. {\bf B287}, 133(1992);

      Z.Phys. {\bf C57}, 349(1993).
      
 \bibitem{varpert} W.E.Caswell, Ann.Phys.(N.Y.){\bf 123}, 153(1979);

      I.G.Halliday and P.Suranyi, Phys.Lett.{\bf B85}, 421(1979);
      
      P.M.Stevenson, Phys.Rev.{\bf D23}, 2916(1981);
      
      J.Killingbeck, J.Phys. {\bf A14}, 1005(1981);
      
      R.Seznec and J.Zinn-Justin, J.Math. Phys. {\bf 20}, 1398(1979);
      
      A.Okopinska, Phys.Rev. {\bf D35}, 1835(1987);
      
      P.M.Stevenson, B.Alles, R.Tarrach, Phys. Rev. {\bf D35}, 2407(1987);
      
      A.Duncan and M.Moshe, Phys.Lett. {\bf B215}, 352(1988);
      
      H.F.Jones and M.Moshe, Phys.Rev. {\bf B234}, 492(1990);
      
      A.Neveu, Nucl.Phys. {\bf B}(Proc. Suppl.)18B, 242(1990);
      
      I.Stancu and P.M.Stevenson, Phys.Rev. {\bf D42}, 2710(1990);
      
      I.Stancu, Phys.Rev. {\bf D43}, 1283(1991);
      
      C.Arvanitis, H.F.Jones and C.S.Parker, Phys.Rev. {\bf D52}, 3704(1995);
      
      W.Janke and H.Kleinert,Phys.Rev.Lett. {\bf 75},2787(1995);
      
      D.Gromes, Z. Phys. {\bf C71}, 347(1996);
      
      C.Arvanitis, F.Geniet and A.Neveu, "Variational solution of the 
      Grosss-Neveu model in the large-N limit", hep-th/9506188;
      
      C.Arvanitis, F.Geniet, M.Iakomi, J.-L.Kneur and A.Neveu,
      "Variational solution of the Gross-Neveu model: Finite N
      and renormalization", hep-th/9511101;
      
      C.Arvanitis, F.Geniet, J.-L.Kneur and A.Neveu, "Chiral
      symmetry breaking in QCD: a variational approach",hep-th/9609247.

 \bibitem{italians} P.Cea, Phys.Lett. {\bf B236}, 191(1989);

      P.Cea and L.Tedesco, "Perturbation theory with a variational
      basis: the generalized Gaussian effective potential", hep-th/
      9607156.
                                   
 \bibitem{lehmann}  G.K\"{a}llen,  Helv. Phys. Acta {\bf 25}, 417(1952);\\
       H.Lehmann, Nuovo Cimento {\bf 11}, 342 (1954).

 \bibitem{chui}  C.K.Chui, "An Introduction to Wavelets", Academic Press,Boston Mass. (1992).

\end{thebibliography}
\end{document}